\documentclass{juliacon}
\usepackage{amsmath}
\begin{document}

% **************GENERATED FILE, DO NOT EDIT**************

\title{Dionysos.jl: a Modular Platform for Smart Symbolic Control}

\author[1]{Julien Calbert}
\author[1]{Adrien Banse}
\author[1]{Benoît Legat}
\author[1]{Raphaël M. Jungers}
\affil[1]{ICTEAM, UCLouvain}


\hypersetup{
pdftitle = {Dionysos.jl: a Modular Platform for Smart Symbolic Control},
pdfsubject = {JuliaCon 2022 Proceedings},
pdfauthor = {Julien Calbert, Adrien Banse, Benoît Legat, Raphaël M. Jungers},
pdfkeywords = {Systems and Control, Symbolic Models, Smart abstractions},
}

\newcommand{\jc}[1]{\textcolor{blue}{[JC]: #1}}
\newcommand{\ab}[1]{\textcolor{green}{[AB:] #1}}
\newcommand{\bl}[1]{\textcolor{red}{[BL]: #1}}
\newcommand{\rj}[1]{\textcolor{purple}{[RJ:] #1}}

\newcommand{\modif}[1]{\textcolor{red}{#1}}

\newtheorem{corollary}{Corollary}
\newtheorem{condition}{Condition}
\newtheorem{assumption}{Assumption}
% \newtheorem{property}{Property}

%%%%%%%%%%%%%%%%%%%%%%%%%%%%%%%%%%%%%%%%%%%%
%%%%%%%%% Classical math symbol %%%%%%%%%%%%
%%%%%%%%%%%%%%%%%%%%%%%%%%%%%%%%%%%%%%%%%%%%
\newcommand{\vect}[1]{#1} % Vector \boldsymbol{#1}
\newcommand{\m}[1]{\mathbf{#1}} % Matrix

\newcommand{\R}{\mathbb{R}}
\newcommand{\N}{\mathbb{N}}
\newcommand{\C}{\mathbb{C}}
\newcommand{\Q}{\mathbb{Q}}
\newcommand{\Z}{\mathbb{Z}}
\newcommand{\abs}[1]{\lvert#1\rvert}
\newcommand{\norm}[1]{\|#1\|}
\newcommand{\defEqual}{\coloneqq}
\newcommand{\mpi}{+} %Moore-Penrose inverse
\newcommand{\argmin}{\operatorname{argmin}} 
\newcommand{\argmax}{\operatorname{argmax}} 
\newcommand{\vol}{\operatorname{Vol}} 
\newcommand{\pdc}[1]{\mathbb{S}_{+}^{#1}} 
\newcommand{\support}[1]{\mathcal{S}(#1)} 
\newcommand{\Int}[1]{\operatorname{int}(#1)} 
\newcommand{\subseteqeq}{\subseteq_0}
\newcommand{\dom}{\operatorname{dom}} 
\newcommand{\sign}{\operatorname{sign}} 
\newcommand{\sev}[1]{\operatorname{sev}\langle #1 \rangle} 
\newcommand{\co}{\operatorname{co}}
\newcommand{\myemptyset}{\varnothing}

%%%%%%%%%%%%%%%%%%%%%%%%%%%%%%%%%%%%%%%%%%%%
%%%%%%%%%% Specific math symbol %%%%%%%%%%%%
%%%%%%%%%%%%%%%%%%%%%%%%%%%%%%%%%%%%%%%%%%%%
\newcommand{\DotSym}{\bullet} 
\newcommand{\ellipsoid}[2]{\operatorname{E}(#1,#2)} 
\newcommand{\ball}[2]{\operatorname{B}(#1,#2)} 
\newcommand{\hyperrectangle}[2]{\operatorname{H}(#1,#2)}

% Systems
\newcommand{\nX}{n_x} 
\newcommand{\nU}{n_u} 
\newcommand{\nW}{n_w} 
\newcommand{\trans}[2][]{\stackrel[#1]{#2}{{\rightsquigarrow}}}
\newcommand{\post}{{\rm Post}} 
\newcommand{\Post}[3]{\post(#1,#3,#2)} %x,u,f
\newcommand{\Available}[1]{\mathcal{U}(#1)}
\newcommand{\set}[1]{\mathcal{#1}}
\newcommand{\Sys}{\mathcal{S}}
\newcommand{\Cont}{\mathcal{C}}
\newcommand{\seq}[1]{\boldsymbol{#1}} % Vector 

% Relation and property 1 
\newcommand{\relOne}{alternating simulation relation} 
\newcommand{\RelOne}{Alternating simulation relation} 
\newcommand{\relOneAbr}{\operatorname{ASR}} 
\newcommand{\propOne}{controlled simulability property} 
\newcommand{\PropOne}{Controlled simulability property} 

% Relation and property 2 
\newcommand{\relTwo}{memoryless concretization relation} 
\newcommand{\RelTwo}{Memoryless concretization relation} 
\newcommand{\relTwoAbr}{\operatorname{MCR}} 
\newcommand{\propTwo}{memoryless concretization property} 
\newcommand{\PropTwo}{Memoryless concretization property} 
\newcommand{\controllerTwo}{memoryless concretized controller} 
\newcommand{\ControllerTwo}{Memoryless concretized controller} 

\newcommand{\ASR}{\operatorname{ASR}}
\newcommand{\FRR}{\operatorname{FRR}} 
\newcommand{\MCR}{\operatorname{MCR}} 

% Relation and property 3 
\newcommand{\relThree}{feedback refinement relation} 
\newcommand{\RelThree}{Feedback refinement relation} 
\newcommand{\relThreeAbr}{\operatorname{FRR}} 
\newcommand{\propThree}{memoryless and symbolic concretization property} 
\newcommand{\PropThree}{Memoryless and symbolic concretization property} 

% Relation and property 4 
\newcommand{\SFR}{\operatorname{SFR}} 

\maketitle
\def\thefootnote{\dagger}\footnotetext{Julien Calbert and Adrien Banse are co-first authors with equal contribution and importance.}\def\thefootnote{\arabic{footnote}}

\begin{abstract}

We introduce \texttt{Dionysos.jl}, a modular package for solving optimal control problems for complex dynamical systems using state-of-the-art and experimental techniques from symbolic control, optimization, and learning.  More often than not with Cyber-Physical systems, the only sensible way of developing a controller is by discretizing the different variables, thus transforming the control task into a purely combinatorial problem on a finite-state mathematical object, called an abstraction of this system. Although this approach offers a safety-critical framework, the available techniques suffer important scalability issues. In order to render these techniques practical, it is necessary to construct smarter abstractions that differ from classical techniques by partitioning the state-space in a non trivial way.

\end{abstract}

\section{Introduction}

%%% INTRO : CPS
In our modern world, the control systems are increasingly complex (think of smart grids, autonomous cars, robots and the internet of things). 
These systems are at the center of a paradigm shift coined as \emph{the cyber-physical revolution} by the industrial and academic communities~\cite{kim2012cyber,alur2015principles,lee2016introduction}.
The industry currently applies classical control techniques even if their requirements are no longer met for these systems. This causes a loss of efficiency and a lack of guarantees that are crucial in view of the importance of safety in such systems~\cite{baier2008principles}. As a consequence, many key technological applications run nowadays at sub-optimal regimes.

\vskip 6pt
%%%  The (mathematical) solution: 
%% abstraction-based control

Formal verification frameworks have been developed using barrier certificates~\cite{prajna2006barrier,prajna2004safety} that system trajectories cannot cross, or using reachability analysis~\cite{althoff2010reachability} to identify potential states that a system can reach.
Reachability analysis involves computing an over-approximation of the set of reachable states, which can be done using the toolbox \texttt{JuliaReach}~\cite{bogomolov2019juliareach}, for example. However, these techniques are mainly focused on verification, and are restricted to some classes of dynamical systems.
In contrast, a renowned approach to address the correct-by-design synthesis relies on \emph{abstractions} (a.k.a. symbolic control)~\cite{tabuada2009verification}, whereby a finite-state machine (also known as ``symbolic model'') approximates the behavior of the original (a.k.a. ``concrete'') system that, instead, evolves in a continuous (or even hybrid) state space. 
This is achieved by defining mathematical relations between the finite state machine and the original dynamics~\cite{alur1998alternating,reissig2016feedback}. 
%% other toolboxes
Several open-source toolboxes~\cite{mazo2010pessoa,Roy2011,rungger2016scots} have already implemented abstraction-based controllers, including recent toolboxes for analysis and synthesis of abstractions for stochastic systems \cite{Mathiesen2024,Wooding2024}, illustrating the need in industry for such techniques offering safety guarantees. Nevertheless, abstraction suffers from the \emph{curse of dimensionality}, and none of these toolboxes can solve non-academic problems of dimension, say, larger than $3$. These limitations of scale explain why abstraction-based techniques have not yet been successful, particularly in the field of robotics.

\vskip 6pt
%%% The (software) solution
To address this issue, we introduce \texttt{Dionysos.jl}\footnote{The package is open source and available on GitHub at: \url{https://github.com/dionysos-dev/Dionysos.jl}.}, a modular Julia \cite{bezanson2017julia} package for solving optimal control problems using state-of-the-art techniques, from control, optimization, and machine learning, for complex systems.
It is built on top of different Julia packages such as \texttt{JuMP.jl} and \texttt{MathOptInterface.jl}, and
features optimal control problem definitions and several abstraction-based methods to solve them.
%% Origin of the project: ERC
\texttt{Dionysos.jl} is the software of the ERC project Learning to control (L2C)%
\footnote{European Reasearch Council (ERC) under the European Union's Horizon 2020 research and innovation program under 
 grant agreement No 864017 - L2C.}.
%
%% New theoretical tools developped in L2C
It implements new fundamental techniques 
that have been designed for smart abstraction~\cite{calbert2021alternating,legat2021abstraction,egidio2022state,banse2023data,calbert2023data} as part of the L2C project, with the goal of designing control techniques that would come with guarantees of safety and efficiency, while at the same time being able to take into account non-standard constraints or information, e.g. coming from first principles in physics, a logical specification translating some legal regulations, or some recommendation from a human being.

\vskip 6pt
In recent years, a few groups have proposed ideas to alleviate the computational burdens of the abstraction-based approach, such as for instance the concept of \emph{lazy abstractions} \cite{camara2011safety,girard2015safety,tazaki2009discrete,hsu2018multi}.
\texttt{Dionysos.jl} provides a general purpose platform allowing to implement such non-standard approaches in a common environment. In particular, it features:
\begin{itemize}
    \item abstractions that are covers of the concrete space (i.e., not only partitions);
    \item discretization templates, such as hyperrectangles and ellipsoids;
    \item controller templates, including piecewise constant controllers or piecewise affine controllers;
    \item different and novel types of abstraction relations such as alternating simulation relation ($\ASR$), feedback refinement relation ($\FRR$), or memoryless concretization relation ($\MCR$).
\end{itemize}

\vskip 6pt
Our techniques, as well as our software solution, have been validated on academic examples from~\cite{girard2009approximately,mouelhi2013cosyma,reissig2016feedback,gol2014language}, accessible in the documentation.

\vskip 6pt
%%% Outline
This paper is structured as follows.  Section~\ref{sec:abstraction} introduces the concept of abstraction-based control. Section~\ref{sec:functionalities} is devoted to the description of the features of \texttt{Dionysos.jl}. Section~\ref{sec:package} describes the package structure and provides a description of its main modules. In Sections~\ref{sec:examples} and~\ref{sec:benchmarks}, we present numerical examples and benchmark comparisons with existing toolboxes, respectively.

\vskip 6pt
\noindent \textbf{Notation:} 
% Intervals
The sets $\R,\Z, \Z_+$ denote respectively the sets of real numbers, integers and non-negative integers.
% Kleene closure
Given a set $A$, the set $A^*$ is its \emph{Kleene closure}. 
% Set-valued map
Given two sets $A,B$, we define a \emph{single-valued map} as $f:A\rightarrow B$, while a~\emph{set-valued map} is defined as $f:A\rightarrow 2^B$, where $2^B$ is the power set of $B$, i.e., the set of all subsets of $B$. 
The image of a subset $\Omega\subseteq A$ under $f:A\rightarrow 2^B$ is denoted $f(\Omega)$.
% Relation
We identify a binary \emph{relation} $R\subseteq A \times B$ with set-valued maps, i.e., $R(a) = \{b \mid (a, b)\in R\}$ and $R^{-1}(b) =\{ a \mid (a, b)\in R\}$.
A relation $R\subseteq A\times B$ is \emph{strict} (resp. \emph{single-valued}) if for every $a\in A$ the set $R(a)\ne \myemptyset$ (resp. $R(a)$ is a singleton). 

\section{Abstraction-based control} \label{sec:abstraction}

In this section, we provide a concise overview of abstraction-based control, the control approach implemented in \texttt{Dionysos.jl}. For a more detailed explanation, please refer to~\cite{tabuada2009verification,reissig2016feedback}.

\subsection{Control framework}

In this section, we start by defining the considered control framework, i.e., the systems, the controllers and the specifications.

\vskip 6pt

We consider dynamical systems of the following form.
\begin{defi}\label{def:sys}
A \emph{transition control system} is a tuple~$\Sys = (\set{X}, \set{U}, F)$ such that
$$x(k+1)\in F(x(k), u(k)),$$
where $\set{X}$ and $\set{U}$ are respectively the set of states and inputs and the set-valued map $F:\set{X}\times \set{U}\rightarrow 2^{\set{X}}$.
\end{defi}

The use of a set-valued map to describe the transition map of a system allows to model perturbations and diverse kinds of non-determinism in a common formalism.
In particular, it can be used to model bounded disturbances $w$, i.e.
\begin{equation}\label{eq:bounded_disturbances}
    F(x,u) = \{f(x,u,w) \mid w\in \set{W}\}
\end{equation}
where $f:\set{X}\times \set{U}\times \set{W}\rightarrow \set{X}$ and $\set{W}\subseteq \R^{\nW}$ is a bounded set.

\vskip 6pt

We introduce the set-valued operator of  \emph{available inputs}, defined as $\mathcal{U}_F(x) = \{ \vect u \in \set{U} \mid F(x,u) \neq \myemptyset \}$, which represents the set of inputs $\vect u$ available at a given state $\vect x$. When it is clear from the context which system it refers to, we simply write the available inputs operator~$\set{U}(x)$.

\vskip 6pt
We say that a transition control system is \emph{deterministic} if for every state $\vect x \in \set{X}$ and control input $\vect u \in \set{U}$, $F(x,u)$ is either empty or a singleton. Otherwise, we say that it is \emph{non-deterministic}. %
A \emph{finite-state} system, in contrast to an \emph{infinite-state} system, refers to a system characterized by finitely many states and inputs.

\vskip 6pt
% Definition of a solution
A tuple $(\seq x, \seq u)\in \set{X}^{T+1}\times \set{U}^{T}$ is a \emph{trajectory} of length $T$ of the system $\Sys=(\set{X},\set{U}, F)$ starting at $x(0)$ if $T\in\N\cup \{\infty\}$, $x(0)\in\set{X}$, $\forall k\in \{0,\ldots,T-1\}: u(k)\in \set{U}(x(k))$ and $x(k+1) \in F(x(k), u(k))$.
The set of trajectories of $\Sys$ is called the \emph{behavior} of $\Sys$, denoted~$\set{B}(\Sys)$.

\vskip 6pt 

We now define \emph{static} controllers, which are characterized by the fact that the set of control inputs that the controller can take is determined solely by the current state of the system.

% Controller's description
\begin{defi}\label{def:controller}
 We define a \emph{static controller} for a system $\Sys = (\set{X},\set{U}, F)$ as a set-valued map $\Cont :\set{X}\rightarrow 2^\set{U}$ such that $\forall \vect x\in\set{X}: \ \Cont(\vect x) \subseteq \set{U}(x)$, $\Cont(x) \ne \myemptyset$. We define the \emph{controlled system}, denoted as $\Cont\times \Sys$, as the transition system characterized by the tuple $(\set{X},\set{U}, F_{\Cont})$ where
 $x'\in F_{\Cont}(x, u) \Leftrightarrow (u\in \Cont(x)\land x'\in F(x,u))$.
\end{defi}

%
%\vskip 6pt
We now define the control problem.
\begin{defi}\label{def:specification}
Consider a system $\Sys = (\set{X},\set{U}, F)$. A \emph{specification} $\Sigma$ for $\Sys$ is defined as any subset $\Sigma \subseteq (\set{X}\times \set{U})^{*} \cup (\set{X}\times \set{U})^{\infty}$. It is said that system $\Sys$ \emph{satisfies} the specification $\Sigma$ if $\set{B}(\Sys)\subseteq \Sigma$. 
A system $\Sys$ together with a specification $\Sigma$ constitute a \emph{control problem} $(\Sys,\Sigma)$.
Additionally, a controller $\Cont$ is said to \emph{solve} the control problem $(\Sys,\Sigma)$ if $\Cont\times \Sys$ satisfies the specification $\Sigma$.
\end{defi}

\subsection{Classical abstraction}

\begin{figure}[ht!]
\centerline{\includegraphics[width=8.5cm]{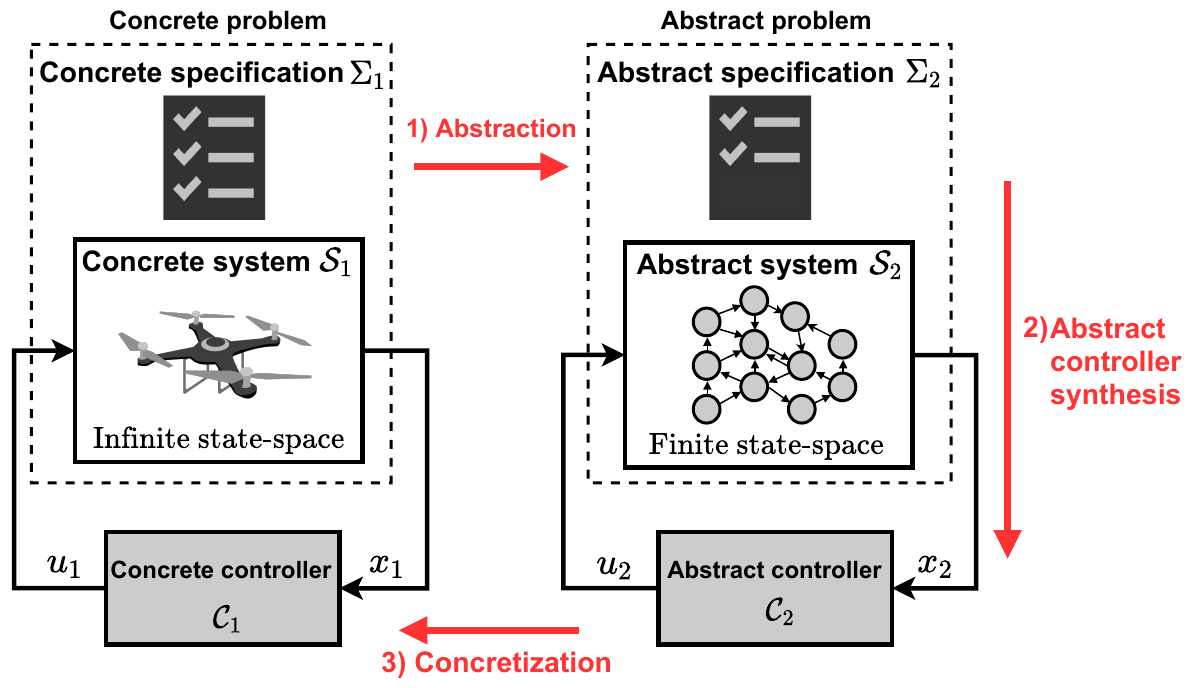}}
\caption{The three steps of abstraction-based control.}
    \label{fig:abstraction-procedure}
\end{figure}

%%% Classical techniques "SCOTS"
Given a mathematical description of the system dynamics and the specifications describing the desired closed-loop behavior of the system, abstraction-based control techniques involve synthesizing a correct-by-construction controller through a systematic three-step procedure illustrated on Figure~\ref{fig:abstraction-procedure}. 
%% Step 1
First, both the original system $\Sys_1=(\set{X}_1,\set{U}_1,F_1)$ and the specifications $\Sigma_1$ are transposed into an \emph{abstract} domain, resulting in an abstract system $\Sys_2=(\set{X}_2,\set{U}_2,F_2)$ and corresponding abstract specifications $\Sigma_2$. We refer to the original system as the \emph{concrete} system as opposed to the abstract system. 
%% Step 2
Next, an abstract controller $\Cont_2$ is synthesized to solve this abstract control problem $(\Sys_2,\Sigma_2)$. 
%% Step 3
Finally, in the third step, called \emph{concretization} as opposed to abstraction, a controller $\Cont_1$ for the original control problem is derived from the abstract controller.

\vskip 6pt
The effectiveness of this approach stems from replacing the concrete system, often characterized by an infinite number of states, with a finite state system. This substitution enables the use of powerful control tools in the second step (see~\cite{belta2017formal,kupferman2001model}), such as those derived from graph theory, including methods like Dijkstra or the A-star algorithm. This facilitates the design of controllers that are correct by construction, often accompanied by rigorous guarantees in terms of safety or performance.

\begin{figure}[ht!]
\centerline{\includegraphics[width=8.5cm]{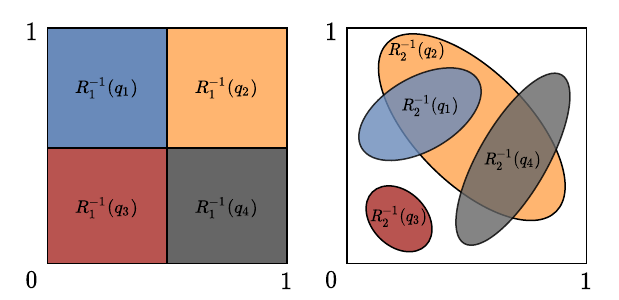}}
\caption{
    Types of discretization of the concrete state space.
    Let $\Sys_{1} = (\set{X}_1,\set{U}_1,F_1)$ with $\set{X}_1 = [0,1]^2$, $\Sys_{2} = (\set{X}_2,\set{U}_2,F_2)$ with $\set{X}_2 = \{q_1,q_2,q_3,q_4\}$, $R_1 \subseteq \set{X}_1\times \set{X}_2$ and $R_2 \subseteq \set{X}_1\times \set{X}_2$ are explicit from the figure.
    Left: $R_1$ is a strict single-valued map, i.e., it induces a full partition of $\set{X}_1$. Right: $R_2$ is a non-strict set-valued map, i.e., it induces a partial cover of $\set{X}_1$.
    }
    \label{fig:discretization}
\end{figure}

\vskip 6pt
In practice, the abstract domain $\set{X}_2$ of $\Sys_2$ is constructed by discretizing the concrete state space $\set{X}_1$ of $\Sys_1$ into subsets (called \emph{cells}). The discretization is induced by a relation $R\subseteq \set{X}_1\times \set{X}_2$, i.e., the cell associated with the abstract state $x_2\in\set{X}_2$ is $R^{-1}(x_2)\subseteq \set{X}_1$.
Note that in this context, we refer to the set-valued map $R(x_1) = \{x_2 \mid (x_1, x_2)\in R\}$ as the \emph{quantizer}.
When $R$ is a single-valued map, we refer to it as defining a \emph{partition} of $\set{X}_1$, in contrast to the case of set-valued maps where we say that it defines a \emph{cover} of $\set{X}_1$, see Figure~\ref{fig:discretization} for clarity.
Notably, the condition that $R$ is a strict relation is equivalent to ensuring that the discretization completely covers $\set{X}_1$.

\vskip 6pt

In order to be able to reconstruct a concrete controller $\Cont_1$ from the abstract controller $\Cont_2$, the relation $R$ must satisfy some properties. It is shown in~\cite[Theorem 1]{Calbert2024a} that if $R$ is an \emph{alternating simulation relation} ($\ASR$)~\cite[Definition 4.19]{tabuada2009verification}, then it is possible to construct a (possibly non-static) controller $\Cont_1$ for $\Sys_1$ from the abstract controller $\Cont_2$.
However, the complexity of the concretization algorithm, and the fact that a non-static controller may have several implementation drawbacks, have motivated researchers to refine the notion of $\ASR$. In particular, most algorithms in the literature~\cite{rungger2016scots,borri2018design,calbert2021alternating} rely on the \emph{feedback refinement relation}~\cite[Def. V.2]{reissig2016feedback}, which we now define.
% \vskip 6pt
\begin{defi}\label{def:FRR}
A strict relation $R$ is a \emph{feedback refinement relation} between the systems $\Sys_1$ and $\Sys_2$, if for each $(x_1,x_2)\in R$
\begin{enumerate}
    \item[(i)] $\set{U}_2(x_2)\subseteq \set{U}_1(x_1)$ \label{cond:i:FRR};
    \item[(ii)] for every $\vect u \in \set{U}_2(x_2),\ \vect x_1'\in F_1(x_1, u), \ \vect x_2'\in R(\vect x_1')$: $ \vect x_2'\in F_2(x_2, u)$.
\end{enumerate}
\end{defi}

This specific relation allows a simple concretization scheme (see~\cite[V.4 Theorem]{reissig2016feedback})
\begin{equation}\label{eq:concrete_controller_FRR}
    \Cont_1(x_1) = \Cont_2(R(x_1)).
\end{equation}
Observe that the requirements $(i)$ and $(ii)$ in Definition~\ref{def:FRR} restrict the class of concrete controllers to piecewise constant controllers since the control input only depends on the abstract state~\eqref{eq:concrete_controller_FRR}. 

\subsection{Smart abstraction} 

\begin{figure}[t]
    \centerline{\includegraphics[width=\linewidth, trim=25 0 0 0, clip]{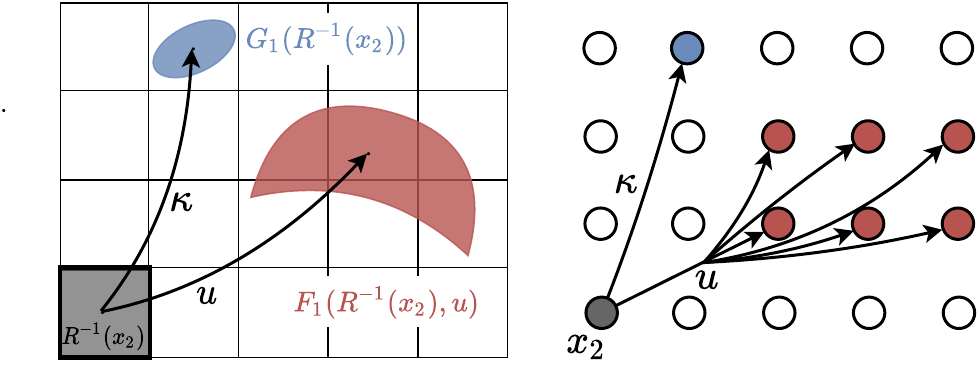}}
    \caption{
    Comparison of piecewise constant and state-dependent controllers.
    The red region illustrates $F_1(x_1, u)$ for all $x_1 \in R^{-1}(x_2)$, given $u \in \set{U}_1$ where $u \in \set{U}_2(x_2)$. The blue region shows $G_1(x_1) = F_1(x_1, \kappa(x_1))$ for all $x_1\in R^{-1}(x_2)$, where $\kappa \in \set{U}_2(x_2)$ is a local state-dependent controller $\kappa:\set{X}_1\rightarrow\set{U}_1$.
    Left: The two-dimensional concrete system with its state space discretization.
    Right: The corresponding abstract system, highlighting the non-deterministic transition $F_2(x_2,u)$ and the deterministic transition $F_2(x_2,\kappa)$.}
	\label{fig:Piece-wise-input}
\end{figure}

The classical abstraction-based approach consists in constructing a feedback refinement relation $R$ based on a predefined partition of the entire state space. 
The limitation to piecewise constant controllers, combined with the use of a rigid/predefined partition of the entire concrete state space, could result in an intractable, or even unsolvable, abstract problem $(\Sys_2, \Sigma_2)$.
Indeed, if the system does not exhibit local incremental stability~\cite[Definition 2.1]{angeli2002lyapunov}\cite{lohmiller1998contraction} around a cell, meaning that trajectories move away from each other, the use of piecewise constant controllers introduces a significant amount of non-determinism into the abstraction as illustrated in Figure~\ref{fig:Piece-wise-input}. 
Indeed, this results in a high cardinality of the set of 
outputs of the transition map $F_2(x_2,u)$ of the abstraction.
\vskip 6pt

%%% Our improvement in Dionysos.jl:
To address these issues, \texttt{Dionysos.jl} provides a framework that generalizes the classical approach by allowing the use of overlapping cells and state-dependent controllers which are defined differently from one cell to another, in a piecewise manner.
The design of low-level controllers within cells, in combination with high-level abstraction-based controllers, opens up new possibilities when co-creating the abstraction and the controller, as is done in so-called \emph{lazy abstractions} (i.e., postponing heavier numerical operations).

\vskip 6pt
%% Use piecewise state-dependent controllers
More precisely, \texttt{Dionysos.jl} allows to construct a (non-strict) relation $R$ that is a cover.
For this purpose, it computes a \emph{memoryless concretization relation} ($\MCR$)~\cite[Definition 8]{Calbert2024a} between $\Sys_1$ and~$\Sys_2$.
\begin{defi}
A relation $R$ is a \emph{memoryless concretization relation} between $\Sys_1$ and $\Sys_2$, if for each $(x_1,x_2)\in R$ 
\begin{align}\label{eq:MCR}
    & \text{for every } \vect u_2\in \set{U}_2(x_2) \text{ there exists } \vect u_1\in \set{U}_1(x_1) \text{ such that }\nonumber\\
    & \hspace{0.5cm}\text{ for every }\vect x_1'\in F_1(x_1,u_1): R(x_1')\subseteq F_2(x_2,u_2).
\end{align}
\end{defi}
We also define the associated \emph{extended relation} $R_e\subseteq \set{X}_1\times \set{X}_2\times \set{U}_1\times \set{U}_2$, which is defined by the set of $(\vect x_1, \vect x_2, \vect u_1, \vect u_2)$ satisfying the condition~\eqref{eq:MCR}, and an \emph{interface} $I_R:\set{X}_1\times \set{X}_2\times \set{U}_2 \rightarrow 2^{\set{U}_1}$ which maps abstract inputs to concrete ones, i.e., 
$$I_R(x_1, x_2, u_2)= \{u_1 \mid (x_1,x_2,u_1,u_2)\in R_e\}.$$

This relation provides a simple concretization scheme, even in the presence of overlapping cells (see~\cite{Calbert2024a} for a complete discussion), which allows the use of state-dependent local controllers within a cell 
$$\Cont_1(x_1) = (\Cont_2\circ_{I_R} R)(x_1) =\bigcup_{x_2\in R(x_1)} I_R(x_1,x_2,\Cont_2(x_2)).$$

By designing these local state-dependent controllers, for example by solving an optimization problem (e.g.,~\cite[Section V]{calbert2024smart}), we can ensure deterministic transitions in the abstraction, thereby eliminating the non-determinism imposed by the discretization of the concrete system (see Figure~\ref{fig:Piece-wise-input}). 
In addition, contrary to the classical approach, this technique avoids discretizing the input-space and uses all the available inputs, making it possible to design, given some cost metric, better control solutions.

\begin{figure}[t]
\centerline{\includegraphics[width=\linewidth, trim=17 0 10 0, clip]{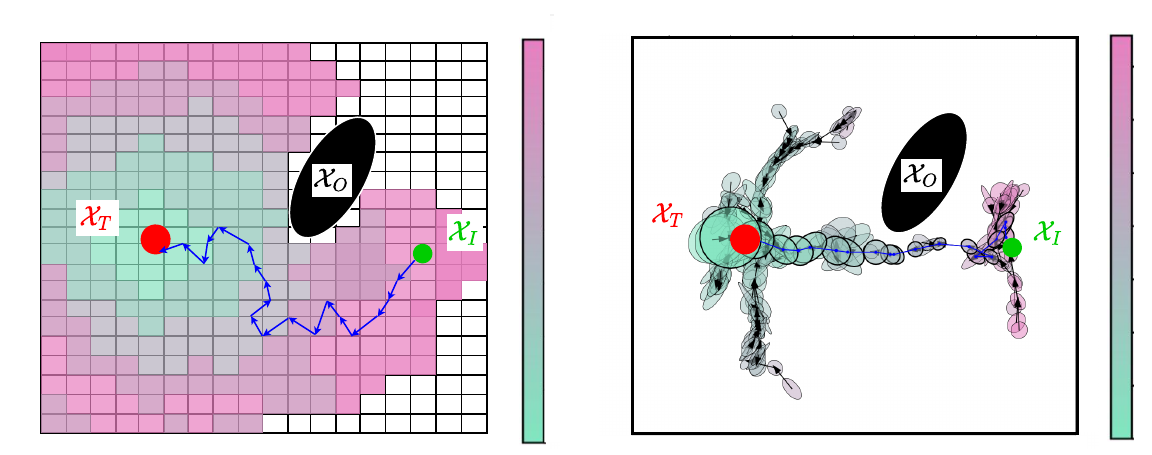}}
\caption{Comparison between classical and smart abstractions for a planar system with state trajectory (blue line) and value function (color map) obtained for the optimal control problem of departing from $\set{X}_I$ and reaching $\set{X}_T$ while avoiding obstacles~$\set{X}_O$. 
Left: Abstraction covering the entire state space with a naive grid-based partition. Non-colored represents a region where no controller could be designed. Right: Abstraction partially covering the state space with ellipsoidal cells and a local feedback controller.}
	\label{fig:Full-vs-Partial}
\end{figure}

\vskip 6pt
%% Partial vs full and standard vs non-standard
In order to reduce the number of cells in the abstraction, \texttt{Dionysos.jl} computes a goal-specific abstraction by co-designing the abstraction and the controller.
This can be done by optimizing the shape of the cells 
during the construction of the abstraction.
For example, the combined use of ellipsoid-based covering and affine local feedback controllers can leverage the power of linear matrix inequalities (or LMI) and convex optimization to create larger/non-standard cells (see Figure~\ref{fig:Full-vs-Partial}).

% \footnotetext{See the documentation~\url{https://dionysos-dev.github.io/Dionysos.jl/stable/generated/Lazy-Ellipsoids-Abstraction/}}

\section{\texttt{Dionysos.jl} features}
\label{sec:functionalities}
In this section, we present the features currently supported by \texttt{Dionysos.jl}.

\vskip 6pt
\textbf{Systems:} 
\texttt{Dionysos.jl} supports bounded disturbances as described in~\eqref{eq:bounded_disturbances} and returns static controllers (see Definition~\ref{def:controller}) for both the abstract and concrete problems.

\vskip 6pt
\textbf{Specifications:}
\texttt{Dionysos.jl} supports either \emph{reach-avoid} or \emph{invariance} (safety) specifications. 
Given a system $\Sys = (\set{X},\set{U},F)$ and sets $\set{X}_I, \set{X}_T, \set{X}_O \subseteq \set{X}$, a \emph{reach-avoid} specification is defined as
\begin{align}
\Sigma^{\text{Reach}} = \{ &(\seq x,\seq u) \in (\set{X} \times \set{U})^\infty \mid x(0) \in \set{X}_I \Rightarrow \nonumber \\
&\exists k \in \Z_+ : \left(x(k) \in \set{X}_T \land \forall k' \in [0;k):x(k')\notin \set{X}_O\right) \}, \label{eq:reach-avoid-spec}
\end{align}
which enforces that all states in the initial set $\set{X}_I$ will reach the target $\set{X}_T$ in finite time while avoiding obstacles in $\set{X}_O$. We use the abbreviated notation $\Sigma^{\text{Reach}}=[\set{X}_I,\set{X}_T,\set{X}_O]$ to denote the specification~\eqref{eq:reach-avoid-spec}. 
Given sets $\set{X}_I,\set{X}_S\subseteq \set{X}$, an \emph{invariance} specification is defined as
\begin{align}
    \Sigma^{\text{safe}} = \{(\seq x,\seq u)\in (\set{X}\times \set{U})^{\infty}\mid\ &x(0)\in \set{X}_I \Rightarrow \nonumber \\
    &\forall k \in \Z_+: x(k)\in \set{X}_S\}, \label{eq:invariance-spec} 
\end{align}
which requires that all states in the initial set $\set{X}_I$ remain (safely) in the set $\set{X}_S$ forever.
In addition to the specifications, a \emph{state cost function} $c:\set{X}\rightarrow \R_+$ evaluating the cost of being in a state $x$, and a \emph{transition cost function} $t:\set{X}\times \set{U}\rightarrow \R_+$ quantifying the cost of transitioning from one state $x$ to another with control input $u$, can be supplied to the control problem. The objective is then to design a controller that satisfies the specification while minimizing the cumulative cost.
\vskip 6pt
\textbf{Discretization templates:} The \emph{quantizer} $R$ can either be a (partial or not) partition or cover of the concrete state space. In addition, given a continuous state space $\set{X}\subseteq \R^n$, \texttt{Dionysos.jl} supports two types of sets for the quantizer: \emph{hyperrectangles} (Definition~\ref{def:hyperrectangle}), and \emph{ellipsoids} (Definition~\ref{def:ellipsoid}).

\begin{defi} \label{def:hyperrectangle}
    A \emph{hyperrectangle} of \emph{center} $c\in\R^n$ and \emph{half-lengths} $h\in \R^n_{+}$ is defined as 
    \[ \hyperrectangle{c}{h} = \{x\in\R^n \mid |x_i-c_i|\le h_i \text{ for }i=1,\ldots,n\}. \]
\end{defi}
\begin{defi} \label{def:ellipsoid}
    An \emph{ellipsoid} with \emph{center} $c\in\R^n$ and \emph{shape} defined by $P\in \pdc{n}$ is defined as 
    \[ \ellipsoid{c}{P} =  \{\vect x\in\R^n\mid (\vect x-\vect c)^\top P(\vect x-\vect c) \le 1\}.\]
\end{defi}

\section{Package structure}
\label{sec:package}

\begin{figure}[t]
    \centerline{\includegraphics[width=5cm]{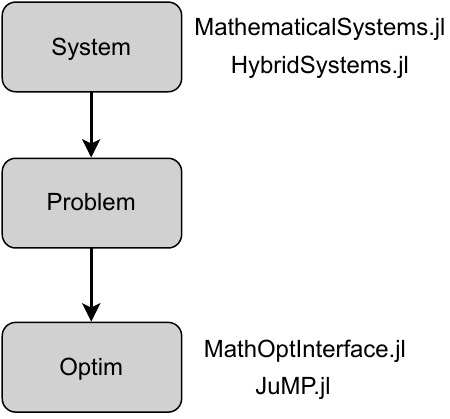}}
    \caption{Summary of Section~\ref{sec:package}. The \texttt{System} module is an extension of \texttt{MathematicalSystems.jl} and \texttt{HybridSystems.jl}, and implements mathematical definitions of dynamical systems. The \texttt{Problem} module contains mathematical definitions of control problems. All problem structures have a system as a field. The \texttt{Optim} module contains the control strategies to solve the problems. It is built on top of the optimization packages \texttt{MathOptInterface.jl} and \texttt{JuMP.jl}.}
    \label{fig:structure}
\end{figure}

In this section we describe the architecture of \texttt{Dionysos.jl}. It is composed of seven root modules. The first three modules, namely \texttt{System}, \texttt{Problem} and \texttt{Optim}, will be described in this paper. For the sake of conciseness, the four other principal modules, namely \texttt{Domain}, \texttt{Mapping}, \texttt{Symbolic} and \texttt{Utils}, are skipped but can be found in the package documentation. A summary of this section is given in Figure~\ref{fig:structure}.

\subsection{The \texttt{System} module}

The \texttt{System} module contains mathematical descriptions of controlled systems (see Definition~\ref{def:sys}), controllers (see Definition~\ref{def:controller}) and trajectories. It is an extension of \texttt{MathematicalSystems.jl} and \texttt{HybridSystems.jl} \cite{BenoitLegat2024}, and complements the latter with more specific system definitions.

\vskip 6pt

The control systems are described as structures implementing the abstract type \texttt{ControlSystem\{N, T\}}, where \texttt{N} and \texttt{T} are respectively the dimension and type of the state-space (e.g. \texttt{N = 3} and \texttt{T = Float64} for a three-dimensional continuous state-space). For example, \texttt{ControlSystemLinearized\{N, T, F1, F2, F3, F4\} <: ControlSystem\{N, T\}}, where \texttt{F1}, \texttt{F2}, \texttt{F3} and \texttt{F4} are subtypes of \texttt{Function}, implements a control system whose transition function has been linearized. It has the form
\begin{equation}
    \dot{x}(t) \in \tilde{F}(x, u), 
\end{equation}
where $\tilde{F}(x, u)$ is such as in \eqref{eq:bounded_disturbances} with an additive noise, that is
\begin{equation}
    \tilde{F}(x, u) = \left\{\tilde{f}(x, u) + w \, | \, w \in \mathcal{W}\right\}, 
\end{equation}
where $\mathcal{W} = [-W, W]^{n_x}$, and where $\tilde{f}$ is the linearized version of some possibly nonlinear function $f$. The system is considered to be sampled with a given time-step, and the corresponding discrete-time transition function is computed with a numerical derivation scheme such as the fourth-order Runge Kutta method, implemented as 
\texttt{RungeKutta4} in \texttt{Dionsysos.jl}. The structure \texttt{ControlSystemLinearized\{N,T,F1,F2,F3,F4\}} contains the fields
\begin{itemize}
    \item \texttt{tstep::Float64}, the sampling time-step,  
    \item \texttt{measnoise::SVector\{N,T\}}, the bound $W$ on the disturbance, 
    \item \texttt{sys\_map::F1}, the sampled possibly nonlinear transition function; 
    \item \texttt{linsys\_map::F2}, the sampled linearized transition function, 
    \item \texttt{error\_map::F3}, the difference between \texttt{linsys\_map} and \texttt{sys\_map}, and
    \item \texttt{sys\_inv\_map::F4}, the inverse of \texttt{sys\_map}.
\end{itemize} 

\vskip 6pt

The controllers implement the abstract type \texttt{Controller} and have a field \texttt{c\_eval} that corresponds to the set-valued function $\Cont(x)$. For example, the structure \texttt{ConstantController\{T,VT\}} implements controllers of the form $\Cont(x) = \{c\}$, where $c$ is a constant. It contains the fields \texttt{c::VT} that is the constant $c$, where \texttt{VT<:AbstractVector\{T\}} and \texttt{T<:Real}, and \texttt{c\_eval}.

\vskip 6pt

Finally, the module \texttt{System} contains descriptions of trajectories. For example, the structure \texttt{ContinuousTrajectory\{T, XVT<:AbstractVector\{T\}, UVT<:AbstractVector\{T\}\}} contains the fields \texttt{x::Vector\{XVT\}} and \texttt{u::Vector\{UVT\}}, and implements trajectories of the form $(\seq x, \seq u)$ such as described in Section~\ref{sec:abstraction}.

\subsection{The \texttt{Problem} module}

The \texttt{Problem} module contains the two structures that respectively define the reach-avoid and invariance specification problems in \texttt{Dionysos.jl}. Both implement the abstract type \texttt{ProblemType}.

\vskip 6pt

\texttt{OptimalControlProblem\{S,XI,XT,XC,TC,T<:Real\}} is the first structure, and implements a reach-avoid problem. Its fields are
\begin{itemize}
    \item \texttt{system::S}, the system to be controlled, 
    \item \texttt{initial\_set::XI}, the initial set $\mathcal{X}_I$, 
    \item \texttt{target\_set::XT}, the target set $\mathcal{X}_T$, 
    \item \texttt{state\_cost::XC}, the state cost function, 
    \item \texttt{transition\_cost::TC}, the transition cost function, and
    \item \texttt{time::T}, the number of allowed steps.
\end{itemize}
Note that, in \texttt{Dionysos.jl}, the obstacles are encoded as part of the domain of the system\footnote{A description of the \texttt{Domain} module can be found in the package documentation.}. The second structure is \texttt{SafetyProblem\{S,XI,XS,T<:Real\}}, and implements an invariance problem. Its fields are 
\begin{itemize}
    \item \texttt{system::S}, the system to be controlled,
    \item \texttt{initial\_set::XI}, the initial set $\mathcal{X}_I$,
    \item \texttt{safe\_set::XS}, the safe set $\mathcal{X}_S$, 
    \item \texttt{time::T}, the number of allowed steps.
\end{itemize}
For both structures, the type \texttt{S} is typically a system type from the packages \texttt{MathematicalSystems.jl} and \texttt{HybridSystems.jl} or from the \texttt{System} module. 

\subsection{The \texttt{Optim} module}

The \texttt{Optim} module contains both abstraction-based and classical control strategies. Table~\ref{tab:solvers} gathers the implemented strategies.
\begin{table}[ht!]
    \centering
    \begin{tabular}{ccc}
        Module & Type & Reference \\ \hline 
        \texttt{BemporadMorari} & Classical & \cite{Bemporad1999} \\
        \texttt{BranchAndBound} & Classical & \cite{legat2021abstraction} \\
        \texttt{AB.UniformGridAbstraction} & Classical abstraction & \cite{reissig2016feedback} \\
        \texttt{AB.EllipsoidsAbstraction} & Smart abstraction & \cite{egidio2022state} \\
        \texttt{AB.HierarchicalAbstraction} & Smart abstraction & \cite{calbert2021alternating} \\
        \texttt{AB.LazyAbstraction} & Smart abstraction & \cite{calbert2021alternating} \\
        \texttt{AB.LazyEllipsoidsAbstraction} & Smart abstraction & \cite{calbert2024smart}.
    \end{tabular}
    \caption{Modules, types and corresponding references of all the control strategies implemented in \texttt{Dionysos.jl}, where \texttt{AB = Abstraction} in the codebase.}
    \label{tab:solvers}
\end{table}
All strategies are viewed as \emph{solvers} inheriting from \texttt{JuMP.jl} \cite{Lubin2023}, a powerful optimization framework embedded in Julia. More precisely, they are all subtypes of the abstract type \texttt{MOI.AbstractOptimizer}, and implement the function \texttt{MOI.optimize!} from the package \texttt{MathOptInterface.jl} \cite{legat2022mathoptinterface}. 

\vskip 6pt

A simple example of an implementation of a classical abstraction-based method is given in the following. In Code~\ref{lst:optimizer_optimize!}, the steps given in Figure~\ref{fig:abstraction-procedure} are followed. In Code~\ref{lst:optimizer_struct}, the structure for the optimizer is defined. To be initialized, it needs a concrete problem to solve, as well as a discretization for the state space and the input space.
\begin{lstlisting}[
    language = Julia, 
    numbers=left, 
    numbersep=6pt, 
    xleftmargin=0.8em,
    label={lst:optimizer_struct}, 
    caption={Definition of an abstraction-based strategy structure implementing the abstract type \texttt{MOI.AbstractOptimizer}.},
    captionpos=b
]
mutable struct Optimizer{T}<:MOI.AbstractOptimizer
    concrete_problem::Union{
        Nothing, 
        PR.OptimalControlProblem, 
        PR.SafetyProblem
    }
    abstract_problem::Union{
        Nothing, 
        PR.OptimalControlProblem, 
        PR.SafetyProblem
    }
    abstract_controller::Any
    concrete_controller::Any
    state_grid::Any
    input_grid::Any 
    function Optimizer{T}(cp,sg,ig) where {T}
        return new{T}(
            cp, 
            nothing, 
            nothing, 
            nothing, 
            nothing, 
            sg, 
            ig
        )
end
end
\end{lstlisting}
\begin{lstlisting}[
    language = Julia, 
    numbers=left, 
    numbersep=6pt, 
    xleftmargin=0.8em,
    label={lst:optimizer_optimize!}, 
    caption={Definition of the \texttt{MOI.optimize!} function for the strategy \texttt{Optimizer} defined in Code~\ref{lst:optimizer_struct}.},
    captionpos=b
]
function MOI.optimize!(opt::Optimizer)
    # Build the abstraction
    abstract_system = build_abs_system(
        opt.concrete_problem.system,
        opt.state_grid,
        opt.input_grid,
    )
    # Build the abstract problem
    abstract_problem = build_abs_problem(
        opt.concrete_problem, 
        abstract_system
    )
    opt.abstract_problem = abstract_problem
    # Solve the abstract problem
    abstract_controller = solve_abs_problem(
        abstract_problem
    )
    opt.abstract_controller = abstract_controller
    # Solve the concrete problem
    opt.concrete_controller = solve_conc_problem(
        abstract_system, 
        abstract_controller
    )
end
\end{lstlisting}

\section{Numerical example}
\label{sec:examples}

In this section, we provide an example of how \texttt{Dionysos.jl} is used to control a nonlinear system with a smart abstraction method. We first define the problem, then solve it with \texttt{Dionysos.jl}, and finally we provide visualization results, as recipes are implemented for all visualizable structures\footnote{See \url{https://docs.juliaplots.org/latest/recipes/} for an introduction on recipes.}. For the sake of conciseness, the code presented in this section is slightly simplified\footnote{See \url{https://dionysos-dev.github.io/Dionysos.jl/stable/generated/Lazy-Ellipsoids-Abstraction/} for the full example.}. 

\vskip 6pt

\begin{exam}\label{ex:ellipsoid}
Let $E(c, P) \subseteq \mathbb{R}^n$ be an ellipsoid with center $c \in \mathbb{R}^n$ and shape defined by $P \in \mathbb{R}^{n \times n}$ as defined in Definition~\ref{def:ellipsoid}.

We consider a reach-avoid problem. The studied dynamical system is noted $(\mathcal{X} \setminus \mathcal{X}_O, \mathcal{U}, F)$, where $\mathcal{X} = [-20, 20]^2$, $\mathcal{X}_O = E(0, 0.02I_2)$, $\mathcal{U} = [-10, 10]^2$, and $F(x, u) = \{f(x, u)\}$. The function $f$ is defined as 
\begin{equation}
    f(x, u) = \begin{pmatrix}
        1.1x_1 - 0.2x_2 - \mu x_2^3 + u_1 \\
        1.1x_2 + 0.2x_1 + \mu x_1^3 + u_2
    \end{pmatrix}, 
\end{equation}
where $\mu = 5 \times 10^{-5}$. We want to solve a reach-avoid problem where $\mathcal{X}_I = E((-10, -10), 10I_2)$ and $\mathcal{X}_T = E((10, 10), I_2)$. There is no state cost, and the transition cost function is defined as 
\begin{equation}
    t(x, u) = x^\top x + u^\top u + 1.
\end{equation}
\end{exam}

\vskip 6pt

The system in Example~\ref{ex:ellipsoid}, as well as other examples, are already defined in \texttt{Dionysos.jl/problems}. The \texttt{OptimalControlProblem} defining the problem in Example~\ref{ex:ellipsoid} can be found in the \texttt{NonLinear} module in \texttt{problems/non\_linear.jl}. In Code~\ref{lst:ex_system}, we import this problem.
\begin{lstlisting}[
    language = Julia, 
    numbers=left, 
    numbersep=6pt,
    xleftmargin=0.8em,
    label={lst:ex_system}, 
    caption={The problem stated above is imported from the \texttt{problems} directory.},
    captionpos=b
]
concrete_problem = NonLinear.problem()
concrete_system = concrete_problem.system
\end{lstlisting}

We choose to use \texttt{AB.LazyEllipsoidsAbstraction} (see Table~\ref{tab:solvers}) to solve this problem, a smart abstraction method that constructs a memoryless concretization relation ($\MCR$) that partially covers the state space with ellipsoids. In Code~\ref{lst:ex_solver}, we instantiate the corresponding \texttt{Optimizer}, then set all the needed fields, including \texttt{concrete\_problem}, the reach-avoid problem stated above. For the sake of conciseness, we do not state the purpose of the other fields, and we gather them in the variable \texttt{other\_parameters} in the code.
\begin{lstlisting}[
    language = Julia, 
    numbers=left, 
    numbersep=6pt, 
    xleftmargin=0.8em,
    label={lst:ex_solver}, 
    caption={The \texttt{Optimizer} is instantiated, and the solver parameters are set.},
    captionpos=b
]
optimizer = MOI.instantiate(
    AB.LazyEllipsoidsAbstraction.Optimizer
)
AB.LazyEllipsoidsAbstraction.set_optimizer!(
    optimizer,
    concrete_problem,
    other_parameters...
)
\end{lstlisting}

In Code~\ref{lst:ex_solve}, we solve the problem and retrieve the corresponding abstract system, abstract problem and concrete controller (see Figure~\ref{fig:abstraction-procedure}). Note that every operation follows the \texttt{MathOptInterface.jl} syntax.
\begin{lstlisting}[
    language = Julia, 
    numbers=left, 
    numbersep=6pt, 
    xleftmargin=0.8em,
    label={lst:ex_solve}, 
    caption={The problem is solved, and the result is extracted from the solver.},
    captionpos=b
]
MOI.optimize!(optimizer)
abstract_system = MOI.get(
  optimizer, 
  MOI.RawOptimizerAttribute("abstract_system")
)
abstract_problem = MOI.get(
  optimizer, 
  MOI.RawOptimizerAttribute("abstract_problem")
)
abstract_controller = MOI.get(
  optimizer, 
  MOI.RawOptimizerAttribute("abstract_controller")
)
concrete_controller = MOI.get(
  optimizer, 
  MOI.RawOptimizerAttribute("concrete_controller")
)
\end{lstlisting}

In Code~\ref{lst:ex_simulate}, we simulate a concrete closed-loop trajectory starting from an initial point $x_0$ sampled from the initial set $\set{X}_I$.
\begin{lstlisting}[
    language = Julia, 
    numbers=left, 
    numbersep=6pt,
    xleftmargin=0.8em,
    label={lst:ex_simulate}, 
    caption={Simulate a closed-loop trajectory.},
    captionpos=b,
]
x0 = UT.sample(concrete_problem.initial_set)
reached(x) = x \in concrete_problem.target_set
trajectory = ST.get_closed_loop_trajectory(
    concrete_system,
    concrete_controller,
    concrete_problem.transition_cost,
    x0;
    stopping = reached,
)
\end{lstlisting}

\begin{figure}
    \centering
    \includegraphics[width=0.8\linewidth]{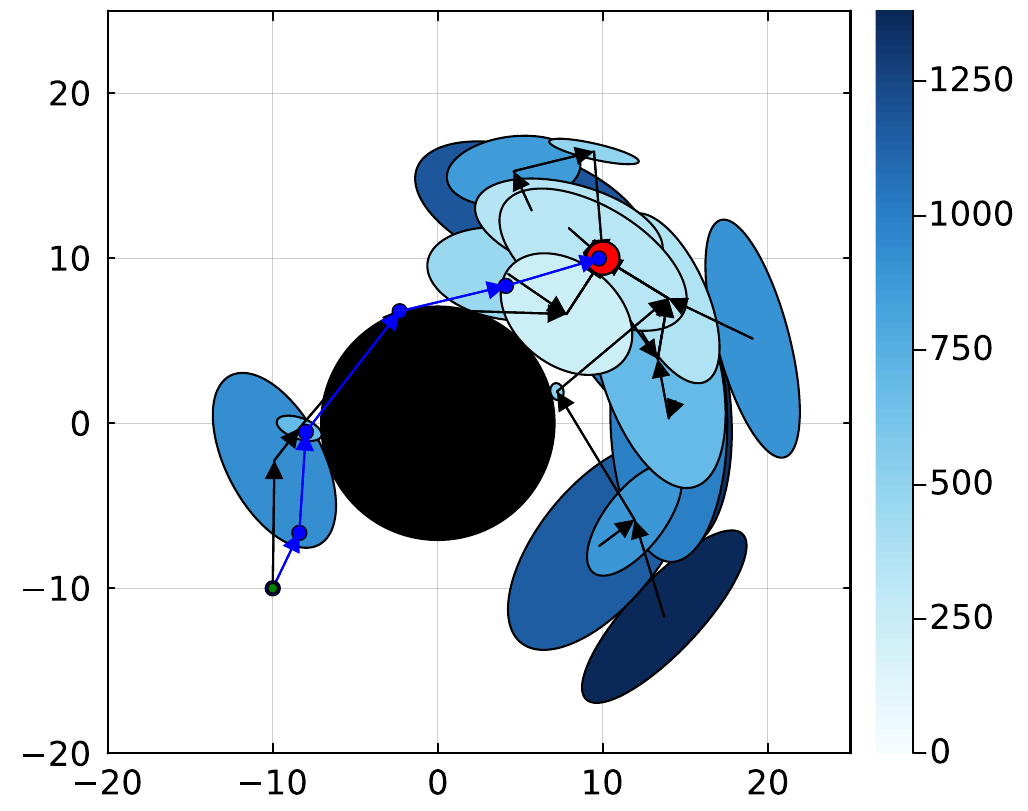}
    \includegraphics[width=0.8\linewidth]{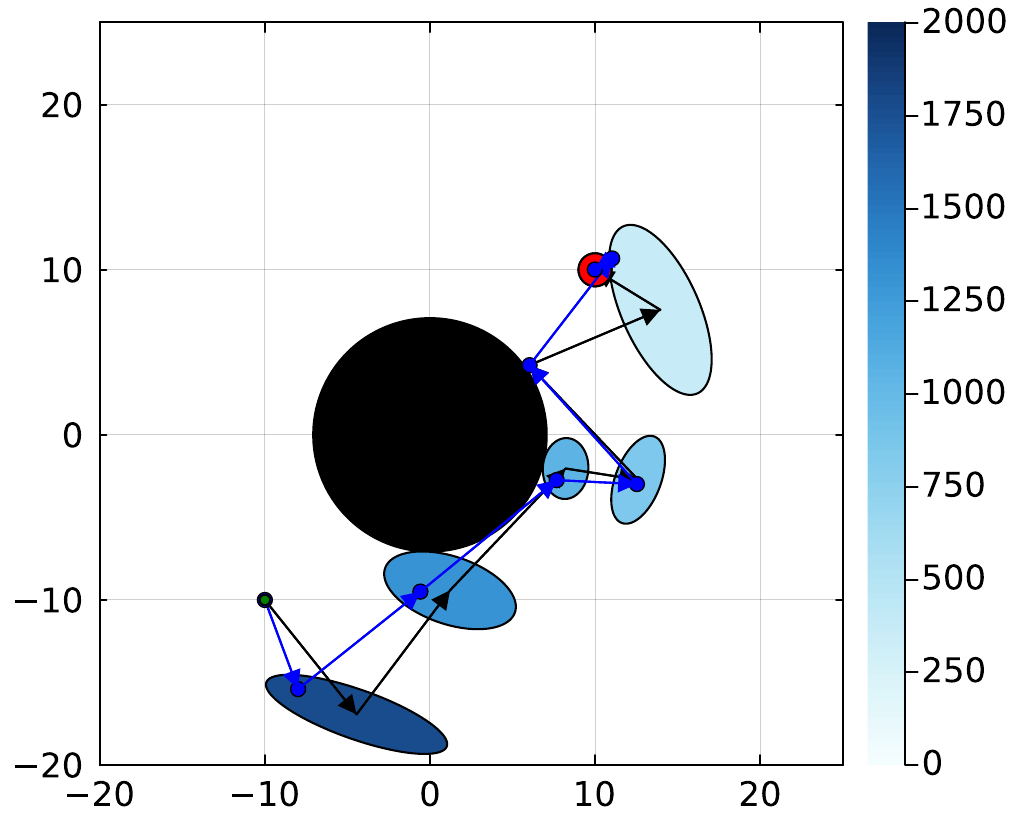}
    \caption{
    Results of Code~\ref{lst:visu} for different meta-parameters (\texttt{other\_parameters}) of the solver \texttt{AB.LazyEllipsoidsAbstraction}. 
    The initial set $\set{X}_I$, target set $\set{X}_T$, and obstacle set $\set{X}_O$ are shown in green, red, and black, respectively. The value function, which provides an upper bound on the cost to reach the target, is shown in a blue color map, with the closed-loop system's trajectory also displayed in blue.}
    \label{fig:ex_plot}
\end{figure}

In \texttt{Dionysos.jl}, we can also generate visualizations thanks to the implemented recipes. In Code~\ref{lst:visu}, we visualize the constructed abstract system, and it gives the plot in Figure~\ref{fig:ex_plot}.
\begin{lstlisting}[
    language = Julia, 
    numbers=left, 
    numbersep=6pt, 
    xleftmargin=0.8em,
    label={lst:visu}, 
    caption={The recipe implemented for the abstraction is used to visualize the abstract system. The \texttt{cost} is set to true to plot the upper-bound on the cost to reach the target.},
    captionpos=b,
]
fig = plot(aspect_ratio=:equal)
plot!(abstract_system; 
      arrowsB = true, 
      cost = true)
plot!(concrete_problem.target_set; 
      color = :red)
plot!(trajectory; 
      color = :blue)
plot!(concrete_problem.initial_set; 
      color = :green)
\end{lstlisting}

% Let us consider the piecewise-affine system $(\mathcal{X}, \mathcal{U}, F)$ with 
% \begin{equation}
%     F(x, u) = \left\{A_{\psi(x)}x + B_{\psi(x)}u + g_{\psi(x)} \right\} \oplus w_{\psi(x)}, 
% \end{equation}
% where $\oplus$ is the Minkowski sum, and $\psi : \mathcal{X} \to \{1, 2, 3\}$ selects one of the 3 subsystems, each of which is associated to a part of $\mathcal{X}$. The three parts $\mathcal{X}_1 = \{x \in \mathcal{X} : x_1 \leq -1\}, \mathcal{X}_3 = \{x \in \mathcal{X} : x_3 > 1\}$, and $\mathcal{X}_2 = \mathcal{X} \setminus (\mathcal{X}_1 \cup \mathcal{X}_3)$. The subsystems are defined by the matrices
% \begin{equation}
% \begin{aligned}
%     A_1 &= \begin{pmatrix}
%             1.01 & 0.3 \\
%             -0.1 & 1.01
%         \end{pmatrix}, &
%     B_1 &= \begin{pmatrix}
%             1 & 0 \\
%             0 & 1
%         \end{pmatrix}, &
%     g_1 &= \begin{pmatrix}
%             -0.1 \\ -0.1
%         \end{pmatrix}, \\
%     A_2 &= A_1^\top, &
%     B_2 &= B_1, &
%     g_2 &= \begin{pmatrix}
%             0 \\ 0
%         \end{pmatrix}, \\
%     A_3 &= A_1, &
%     B_3 &= B_1, &
%     g_3 &= -g_1, 
% \end{aligned}
% \end{equation}
% and the additive noises are $w_1 = w_2 = w_3 = [-0.05, 0.05]^2$. The state space is $\mathcal{X} = [-2, 2]^2$ and the input space is $\mathcal{U} = [-0.5, 0.5]^2$. The goal is 

\section{Benchmarking}
\label{sec:benchmarks}

In order to evaluate the performance of \texttt{Dionysos.jl}, we compare the performance of our package against other similar packages, namely SCOTS \cite{rungger2016scots} and CoSyMA \cite{mouelhi2013cosyma}. We exclude PESSOA \cite{mazo2010pessoa, Roy2011} from the comparison as it is outperformed by SCOTS \cite{rungger2016scots}. The code for comparing these packages is published on CodeOcean \cite{Calbert2024}, and is entirely reproducible. For more information, we invite the reader to read the \texttt{README.md} file in the CodeOcean capsule.

\vskip 6pt

We reproduced the two numerical experiments of \cite{rungger2016scots}. First, the DC-DC converter example presented in \cite[Section 4.2]{rungger2016scots} is reproduced with \texttt{Dionysos.jl}, SCOTS and CoSyMA. Thanks to the modularity of \texttt{Dionysos.jl}, we can specify to the package that the system is incrementally stable, resulting in sped-up abstraction and synthesis procedures \cite{Camara2011}. For the sake of completeness, we also provide the performance of \texttt{Dionysos.jl} in the setting where no prior knowledge on the stability is given. The results can be found in Table~\ref{tab:dcdc_bench}.

\begin{table}[ht!]
    \centering
    \begin{tabular}{c|ccc}
        & Abstraction [s] & Synthesis [s] & Total [s] \\ 
        \hline 
        \texttt{Dionysos.jl} (no prior) & \textbf{1.24} & \textbf{3.53} & \textbf{4.77} \\
        \texttt{Dionysos.jl} (prior) & \textbf{0.63} & \textbf{2.76} & \textbf{3.39} \\
        SCOTS & 19.05 & 74.01 & 93.06 \\
        CoSyMA & --- & --- & 5.31
    \end{tabular}
    \caption{Comparison between SCOTS, CoSyMA and \texttt{Dionysos.jl} (\texttt{AB.UniformGridAbstraction} solver from Table~\ref{tab:solvers}) for the DC-DC converter example. \texttt{Dionysos.jl} outperforms SCOTS and CoSyMA with and without prior knowledge of the system's incrementally stable property.}
    \label{tab:dcdc_bench}
\end{table}

Second, the path planning problem presented in \cite[Section 4.1]{rungger2016scots} is executed with \texttt{Dionysos.jl} and SCOTS. We exclude CoSyMA because this system is not incrementally stable. The results can be found in Table~\ref{tab:vehicle_bench}.

\begin{table}[ht!]
    \centering
    \begin{tabular}{c|ccc}
        & Abstraction [s] & Synthesis [s] & Total [s] \\ 
        \hline 
        \texttt{Dionysos.jl} & \textbf{8.58} & \textbf{6.45} & \textbf{15.03} \\
        SCOTS & 117.52 & 480.44 & 597.96 \\
    \end{tabular}
    \caption{Comparison between SCOTS, CoSyMA and \texttt{Dionysos.jl} (\texttt{AB.UniformGridAbstraction} solver from Table~\ref{tab:solvers}) for the path planning example. \texttt{Dionysos.jl} outperforms SCOTS.}
    \label{tab:vehicle_bench}
\end{table}

\begin{figure}[ht!]
    \centerline{\includegraphics[width=\linewidth, trim=90 0 85 0, clip]{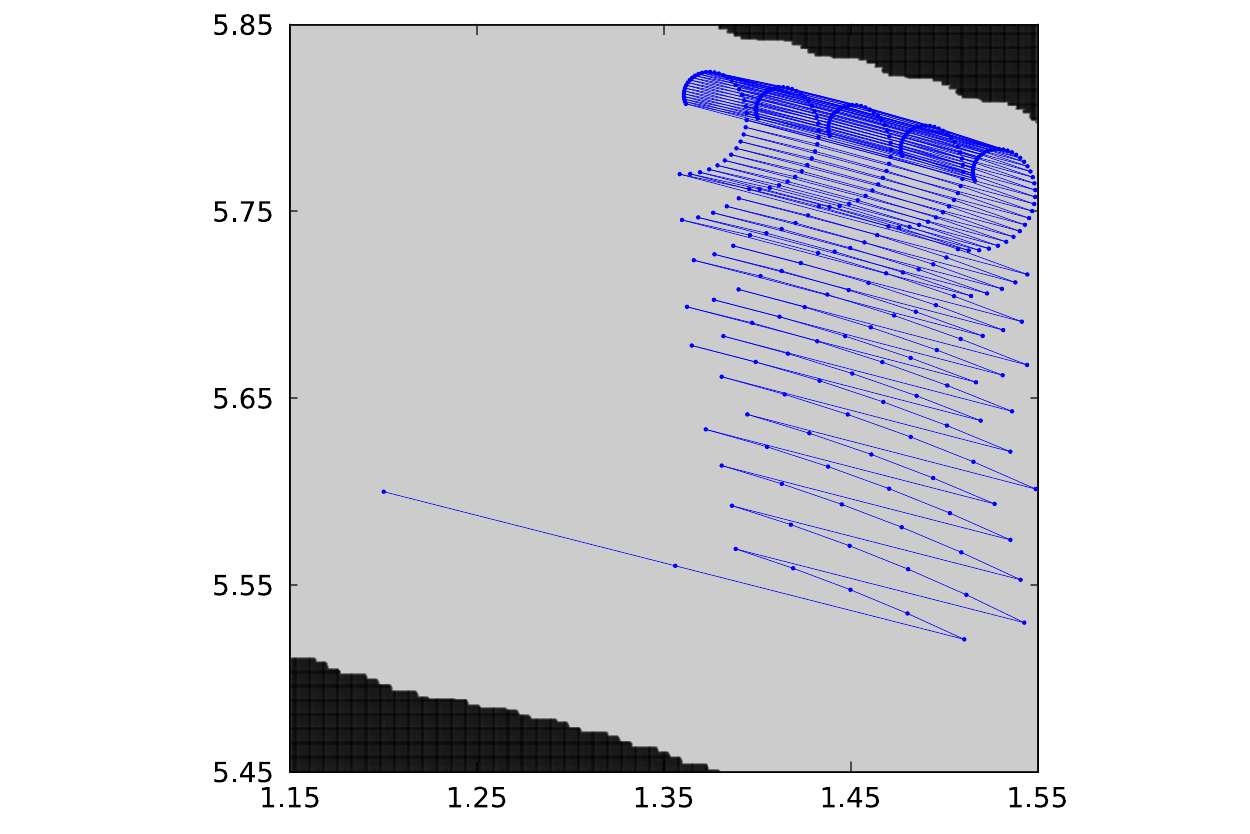}}
    \caption{Reproduction of \cite[Figure 4]{rungger2016scots} with \texttt{Dionysos.jl}.}
    \label{fig:dcdc}
\end{figure}

\begin{figure}[ht!]
    \centerline{\includegraphics[width=\linewidth, trim=90 0 85 0, clip]{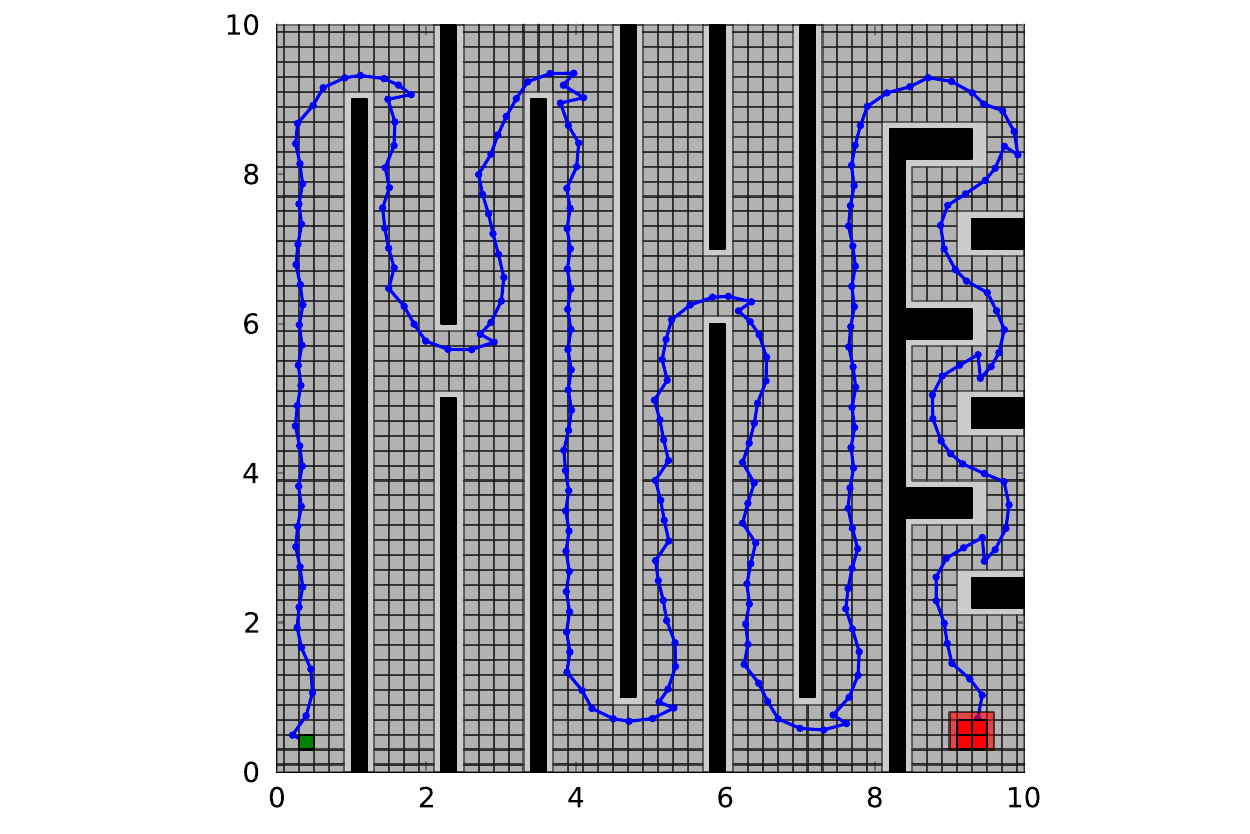}}
    \caption{Reproduction of \cite[Figure 3]{rungger2016scots} with \texttt{Dionysos.jl}.}
    \label{fig:vehicle}
\end{figure}

We see that \texttt{Dionysos.jl} outperforms the other packages for these two examples. We also reproduced \cite[Figures 3 and 4]{rungger2016scots} in Figure~\ref{fig:dcdc} and Figure~\ref{fig:vehicle}, which shows that they compute the same controller. The visualizations of the DC-DC converter controller with and without prior knwoledge are identical, as it can be verified in \cite{Calbert2024}.

\vskip 6pt

The reason for such a difference between SCOTS, written in C++, and \texttt{Dionysos.jl} does not lie in the programming language used to write the package but in the synthesis algorithm itself. For example, unlike SCOTS, our package does not make use of \emph{Binary Decision Diagrams} (or \emph{BDDs})~\cite{bryant1992symbolic}, which as recognized in~\cite{rungger2016scots} results in substantially longer execution times compared to tools that use alternative data structures.

 
\section{Conclusions and further work}

%%% Summary
In this paper, we introduce \texttt{Dionysos.jl}, a new software package that provides both a new abstract symbolic representation of the system and controllers with safety guarantees. It generalizes existing toolboxes limited to classical abstractions by allowing the construction of abstractions within the memoryless concretization relation framework, which provides a simple concretization step and the design of low-level controllers.
We provide a description of the structure of the package, and describe further the main modules of it. We then show how \texttt{Dionysos.jl} can be used in practice by providing a reach-avoid control problem example. Finally, we demonstrated the performance of our package compared to existing similar toolboxes.

\vskip 6pt
As outlined in Section~\ref{sec:package} with the array of implemented solvers, the goal of \texttt{Dionysos.jl} is to provide a modular environment to facilitate the implementation of new smart abstraction algorithms based for instance on a partial cover of the state-space and the use of piecewise state-dependent controllers.

\vskip 6pt
%%% Future work
Because of the hardness of the control problem we aim to solve, the choice of an appropriate solver and its meta-parameters can be self-tuned by end-users.
In future work, we plan to design a meta-solver within \texttt{Dionysos.jl} which would combine these modules in an ad-hoc and opportunistic approach, thanks to Machine Learning and Artificial Intelligence techniques, in order to exploit the particular problem structures and alleviate the curse of dimensionality.

\section{Acknowledgment}
JC and AB are FRIA Research Fellows.
The research of BL is supported by the European Commission (ERC Adv. Grant) under Grant 885682.
RJ is a FNRS honorary Research Associate. This project has received funding from the European Research Council (ERC) under the European Union's Horizon 2020 research and innovation programme under grant agreement No 864017 - L2C.

%
% **************GENERATED FILE, DO NOT EDIT**************

\bibliographystyle{juliacon}
\bibliography{ref.bib}

\end{document}